\documentclass[12pt]{article}
\usepackage{graphicx}
\usepackage{lscape}
\usepackage{citesort}
\usepackage{amssymb}
\usepackage{appendix}
\usepackage{multirow}

\begin{document}
\def\qq{\langle \bar q q \rangle}
\def\uu{\langle \bar u u \rangle}
\def\dd{\langle \bar d d \rangle}
\def\sp{\langle \bar s s \rangle}
\def\GG{\langle g_s^2 G^2 \rangle}
\def\Tr{\mbox{Tr}}
\def\figt#1#2#3{
        \begin{figure}
        $\left. \right.$
        \vspace*{-2cm}
        \begin{center}
        \includegraphics[width=10cm]{#1}
        \end{center}
        \vspace*{-0.2cm}
        \caption{#3}
        \label{#2}
        \end{figure}
	}
	
\def\figb#1#2#3{
        \begin{figure}
        $\left. \right.$
        \vspace*{-1cm}
        \begin{center}
        \includegraphics[width=10cm]{#1}
        \end{center}
        \vspace*{-0.2cm}
        \caption{#3}
        \label{#2}
        \end{figure}
                }

\def\ds{\displaystyle}
\def\beq{\begin{equation}}
\def\eeq{\end{equation}}
\def\bea{\begin{eqnarray}}
\def\eea{\end{eqnarray}}
\def\beeq{\begin{eqnarray}}
\def\eeeq{\end{eqnarray}}
\def\ve{\vert}
\def\vel{\left|}
\def\ver{\right|}
\def\nnb{\nonumber}
\def\ga{\left(}
\def\dr{\right)}
\def\aga{\left\{}
\def\adr{\right\}}
\def\lla{\left<}
\def\rra{\right>}
\def\rar{\rightarrow}
\def\lrar{\leftrightarrow}  
\def\nnb{\nonumber}
\def\la{\langle}
\def\ra{\rangle}
\def\ba{\begin{array}}
\def\ea{\end{array}}
\def\tr{\mbox{Tr}}
\def\ssp{{\Sigma^{*+}}}
\def\sso{{\Sigma^{*0}}}
\def\ssm{{\Sigma^{*-}}}
\def\xis0{{\Xi^{*0}}}
\def\xism{{\Xi^{*-}}}
\def\qs{\la \bar s s \ra}
\def\qu{\la \bar u u \ra}
\def\qd{\la \bar d d \ra}
\def\qq{\la \bar q q \ra}
\def\gGgG{\la g^2 G^2 \ra}
\def\q{\gamma_5 \not\!q}
\def\x{\gamma_5 \not\!x}
\def\g5{\gamma_5}
\def\sb{S_Q^{cf}}
\def\sd{S_d^{be}}
\def\su{S_u^{ad}}
\def\sbp{{S}_Q^{'cf}}
\def\sdp{{S}_d^{'be}}
\def\sup{{S}_u^{'ad}}
\def\ssp{{S}_s^{'??}}

\def\sig{\sigma_{\mu \nu} \gamma_5 p^\mu q^\nu}
\def\fo{f_0(\frac{s_0}{M^2})}
\def\ffi{f_1(\frac{s_0}{M^2})}
\def\fii{f_2(\frac{s_0}{M^2})}
\def\O{{\cal O}}
\def\sl{{\Sigma^0 \Lambda}}
\def\es{\!\!\! &=& \!\!\!}
\def\ap{\!\!\! &\approx& \!\!\!}
\def\ar{&+& \!\!\!}
\def\ek{&-& \!\!\!}
\def\kek{\!\!\!&-& \!\!\!}
\def\cp{&\times& \!\!\!}
\def\se{\!\!\! &\simeq& \!\!\!}
\def\eqv{&\equiv& \!\!\!}
\def\kpm{&\pm& \!\!\!}
\def\kmp{&\mp& \!\!\!}
\def\mcdot{\!\cdot\!}
\def\erar{&\rightarrow&}

% .........................................................

\def\simlt{\stackrel{<}{{}_\sim}}
\def\simgt{\stackrel{>}{{}_\sim}}

\def\olra{\stackrel{\leftrightarrow}}
\def\ola{\stackrel{\leftarrow}}
\def\ora{\stackrel{\rightarrow}}

% .........................................................

\title{
         {\Large
                 {\bf
Strong decay constants of heavy tensor mesons in light cone QCD sum rules
                 }
         }
      }

\author{\vspace{1cm}\\
{\small
H. A. Alhendi$^{1,}$\thanks {e-mail:
alhendi@KSU.EDU.SA}\,\,,
\small T. M. Aliev$^{2,}$\thanks {e-mail:
taliev@metu.edu.tr}~\footnote{permanent address:Institute of
Physics,Baku,Azerbaijan}\,\,,
\small M. Savc{\i}$^{2,}$\thanks
{e-mail: savci@metu.edu.tr}} \\
{\small $^1$ Physics and Astronomy Department, King Saud University, Saudi
Arabia}\\
{\small $^2$ Physics Department, Middle East Technical University,
06531 Ankara, Turkey }
}

\date{}

\begin{titlepage}
\maketitle
\thispagestyle{empty}

\begin{abstract}

Strong decay constants of the heavy tensor to heavy pseudoscalar
(vector) and light pseudoscalar mesons are estimated within the light cone
QCD sum rules. It is obtained that the values of these coupling constants
are very strongly dependent on the choice of the Lorentz structure.
Moreover, using these strong coupling constants the corresponding decay
widths are calculated.
 
\end{abstract}

\vspace{1cm}
~~~PACS number(s): 11.55.Hx, 13.20.He, 13.40.Em
\end{titlepage}

\section{Introduction}

Last decade was very fruitful in spectroscopy of the rapidly growing number
of particles. Many new particles are discovered (see
\cite{Rozd01,Rozd02,Rozd03,Rozd04,Rozd05,Rozd06,Rozd07,Rozd08,Rozd09}
and the references therein) which part of them are successfully
described in framework of the quark model, and interpretation of the
remaining ones demand going beyond the quark picture. In particular, the
masses and decay widths of the excited mesons such as
${\cal D}_2^\ast(2400)$, ${\cal D}_{S_2}^\ast(2573)$,
${\cal B}_2^\ast(5747)$ and ${\cal B}_{S_2}^\ast(5840)$ with the quantum
number $J^P=2^+$ are measured in experiments \cite{Rozd08,Rozd09,Rozd10}.
The properties of ${\cal D}$--wave and radial excited heavy light meson
system will be examined in detail in future planned experiments at LHBc and
KEK--B.

The strong decays of ${\cal D}$--wave mesons, such as
${\cal D}_2 (2460) \to {\cal D}^{\ast +} \pi^-,~{\cal D}^+ \pi^-$,
${\cal D}_2^+ (2460) \to {\cal D}^0 \pi^+$
\cite{Rozd06,Rozd12,Rozd13,Rozd14}, ${\cal D}_{S_2}^+ (2573) \to {\cal D}^0
K^+$ \cite{Rozd06}, ${\cal B}_2 (5747) \to {\cal B}^{\ast +} \pi^-,~{\cal
B}^+ \pi^-$, \cite{Rozd09,Rozd10}, and ${\cal B}_{S_2}^+ (5840) \to {\cal
B}^+ K^-$ \cite{Rozd09,Rozd10} have already been observed in experiments.
The decay constants of the tensor mesons ${\cal D}_2^\ast (2460)$ and
${\cal D}_{S_2}^\ast (2573)$ have been studied within the three--point QCD
sum rules method in \cite{Rozd15}. In framework of the same approach the
strong constants of ${\cal D}_2^\ast (2460) \to {\cal D} \pi$,
${\cal D}_{S_2}^\ast (2573) \to {\cal D} K$, ${\cal B}_2^\ast (5747) \to
{\cal B} \pi$, and ${\cal B}_{S_2}^\ast (5840) \to {\cal B} K$ 
transitions have been calculated in \cite{Rozd16}.
Recently, the strong decay constants of
the ${\cal D}_2^\ast (2460) \to {\cal D}^\ast \pi$, ${\cal D}_{S_2}^\ast
(2573) \to {\cal D}^\ast K$, ${\cal B}_2^\ast (5747) \to    
{\cal B}^\ast \pi$, and ${\cal B}_{S_2}^\ast (5840) \to {\cal B}^\ast K$
transitions have been studied in \cite{Rozd17} in framework of the
three-point QCD sum rules and the local QCD methods.

The present work is devoted to the calculation of the constants of the
tensor mesons ${\cal D}_2^\ast (2460)$, ${\cal D}_{S_2}^\ast
(2573)$, ${\cal B}_2^\ast (5747)$, and ${\cal B}_{S_2}^\ast (5840)$ in
framework of the QCD sum rules method.

The paper is arranges as follows. In section 2, the light cone QCD sum
rules derived for the coupling constants of the ${\cal D}_2^\ast \to
{\cal D} \pi$, ${\cal D}_{S_2}^\ast
\to {\cal D} K$, ${\cal B}_2^\ast \to    
{\cal B} \pi$, and ${\cal B}_{S_2}^\ast \to {\cal B} K$
transitions. Section 2 is devoted to the numerical analysis of the
aforementioned coupling constants, and present the values of the
corresponding two--body strong decays. In this section, comparison of our
results with the ones existing literature, and our concluding remarks
are also presented.  

\section{Light cone QCD sum rules for the strong coupling constants of heavy
tensor mesons}

In this section we derive the light cone QCD sum rules for the strong
coupling constants of heavy tensor mesons with heavy pseudoscalar (vector)
mesons with participation of the light pseudoscalar mesons. For this aim we
consider the following correlation function,

\bea
\label{eozd01}
\Pi_{\mu\nu~(\tau)} (p,q) = i \int d^4x e^{ipx} \lla {\cal P}(q) \vel
J_{\mu\nu~(\tau)} (x) J_{5(\tau)}^\dag (0) \ver 0 \rra~,
\eea
where
\bea
\label{nolabel01}
J_{\mu\nu} (x) = {1\over 2} \Big[ \bar{q} (x) \gamma_\mu \olra{\cal D}_\nu Q (x) + 
 \bar{q} (x) \gamma_\nu \olra{\cal D}_\mu Q (x) \Big]~,\nnb
\eea
is the interpolating current of the heavy tensor meson, $J_5=\bar{q} i
\gamma_5 Q$, and $J_\tau= \bar{q} \gamma_\tau Q$ are the interpolating
currents of the heavy pseudoscalar and vector mesons, respectively; ${\cal
P}$
denotes the light pseudoscalar meson, $q$ and $Q$ in the expressions of the
interpolating currents describes the light and heavy quarks. The covariant
derivative $\olra{\cal D}_\mu$ is defined in the following way,
\bea
\label{eozd02}
\olra{\cal D}_\mu (x) \es {1\over 2} \Big[
\ora{\cal D}_\mu (x) -
\ola{\cal D}_\mu (x) \Big]~, \mbox{and,}\\
\label{eozd03}
\ora{\cal D}_\mu (x) \es \ora{\partial}_\mu (x) - i
{g\over 2} \lambda^a A_\mu^a (x) ~, \nnb \\
\ola{\cal D}_\mu (x) \es \ola{\partial}_\mu (x) + i
{g\over 2} \lambda^a A_\mu^a (x)~,
\eea
where $\lambda^a$ are the Gell-Mann matrices, and $A_\mu^a (x)$ is the
external gluon field, and in our calculations we shall use the
Fock--Schwinger gauge, i.e., $x_\mu A^\mu (x) = 0$. The advantage of this
gauge is that the gluon field can be expressed in terms of gluon field
strength tensor define as,
\bea
\label{eozd04}
A_\mu^a \es \int_0^1 dt\, G_{\nu\mu} (tx) x^\nu \nnb \\
\es {1\over 2} x^\nu G_{\nu\mu} (0) + {1\over 3} x^\alpha x^\nu {\cal
D}_\alpha G_{\nu\mu} + \cdots~.
\eea

In order to construct the sum rules for the strong coupling constants of the
heavy tensor mesons with heavy pseudoscalar (vector) mesons with the
participation light pseudoscalar mesons, we shall calculate the correlation
function (\ref{eozd01}) in two different ways: a) In terms of hadronic
parameters; b) in terms of quarks and gluons. Equating then these two
representations, we derive the sum rule for the aforementioned hadronic
coupling constant.

We first consider calculation of the correlator function from the hadronic
side. Inserting a complete set of hadrons carrying the same quantum numbers
as the interpolating currents $J_{\mu\nu}$ and $J_{5(\tau)}$ into the
correlation function, and isolating the ground state contributions of the
heavy tensor and heavy pseudoscalar (vector) mesons we get,
\bea
\label{eozd05} 
\Pi_{\mu\nu(\tau)} = {\lla {\cal P} (q) T_Q \ve {\cal P}_Q (V_Q) \rra \lla 0 \vel
J_{\mu\nu} \ver T_Q \rra \lla P_Q (V_Q) \vel J_{5(\tau)}^\dag \ver 0 \rra   
\over (p^2-m_{T_Q}^2) ( p^{\prime 2} - m_{P_Q(V_Q)}^2)}+ \cdots ~,
\eea
where $T_Q,P_Q,V_Q,{\cal P}$ mean heavy tensor, heavy pseudoscalar, heavy vector
mesons, and light pseudoscalar mesons, respectively,
and dots mean higher states and continuum contributions.

The matrix elements in Eq. (\ref{eozd05}) are defined as,
\bea
\label{eozd06}
\lla 0 \vel J_{\mu\nu} \ver T_Q(p) \rra \es f_T m_{T_Q}^3 \epsilon_{\mu\nu}^\ast (s,p)~,\\
\label{eozd07}
\lla 0 \vel J_5 \ver P_Q(p) \rra \es {f_Q m_{P_Q}^2 \over m_Q+m_q}~,\\
\label{eozd08}
\lla 0 \vel J_\tau \ver V_Q(p) \rra \es f_V m_{V_Q}
\xi_\tau^\ast(p^\prime)~,\\
\label{eozd09}
\lla {\cal P} T_Q \ve P_Q\rra \es g_1 \epsilon_{\alpha\beta} p^{\prime\alpha}
p^{\prime\beta} = g_1 \epsilon_{\alpha\beta} q^\alpha q^\beta~,\\
\label{eozd10}
\lla {\cal P} T_Q \ve V_Q\rra \es g_2 \varepsilon^{\alpha\beta
\eta\lambda} p_\alpha \epsilon_{\beta\varphi} q^\varphi
p_\eta^\prime \xi_\lambda~,
\eea
where $\epsilon_{\mu\nu}$ is the polarization vector of the heavy tensor
meson, $f_T$, $f_Q$, $f_V$ are the decay constants of the tensor, heavy
pseudoscalar, and heavy vector mesons; $g_1$ and $g_2$ are the hadronic
coupling constants, and $s$ is the polarization vector of the heavy vector
mesons.
Performing summation over the
spins of the heavy tensor and heavy vector mesons in Eqs.
(\ref{eozd05}--\ref{eozd10}), we get for the correlation
functions describing the heavy tensor mesons with heavy pseudoscalar,
heavy vector mesons and light pseudoscalar mesons, the following results,
respectively,
\bea
\label{eozd11}
\Pi_{\mu\nu} \es g_1 {f_T m_{T_Q}^3 \over p^2-m_{T_Q}^2}
{f_Q m_{P_Q}^2 \over m_Q+m_q}
{q^\alpha q^\beta \over p^{\prime 2} - m_{P_Q}^2}
\Bigg\{ {1\over 2} \Big[ \tilde{g}_{\mu\alpha} \tilde{g}_{\nu\beta} +
\tilde{g}_{\mu\beta} \tilde{g}_{\nu\alpha} \Big] - {1\over 3}
\tilde{g}_{\mu\nu} \tilde{g}_{\alpha\beta} \Bigg\}~,\nnb\\
\es g_1 {f_T m_{T_Q}^3 \over p^2-m_{T_Q}^2}
{f_Q m_{P_Q}^2 \over m_Q+m_q}
{1 \over p^{\prime 2} - m_{P_Q}^2} \Bigg\{g_{\mu\nu}
\Bigg( - {1\over 3} m_{\cal P}^2 + {(m_{T_Q}^2 + m_{\cal P}^2 - m_{P_Q}^2)^2 \over
12 m_{T_Q}^2} \Bigg) + q_\mu q_\nu \nnb \\
\ek {1\over 2} (q_\mu p_\nu +
q_\nu p_\mu) {(m_{T_Q}^2 + m_{\cal P}^2 - m_{P_Q}^2) \over 
m_{T_Q}^2} \nnb \\
\ar {p_\mu p_\nu \over m_{T_Q}^4}
\Bigg[ {(m_{T_Q}^2 + m_{\cal P}^2 - m_{P_Q}^2)^2 \over 6} +
{m_{\cal P}^2 m_{T_Q}^2 \over 3} \Bigg] \Bigg\}~, \\ \nnb \\    
\label{eozd12}
\Pi_{\mu\nu\tau} \es g_2 f_V m_{V_Q}
{f_T m_{T_Q}^3 \over p^2-m_{T_Q}^2} {1\over p^{\prime 2} - m_{V_Q}^2}
\varepsilon^{\alpha\beta\eta\lambda} p_\alpha q^\varphi p_\eta^\prime
\Bigg\{ {1\over 2} \Big[\tilde{g}_{\mu\beta} \tilde{g}_{\nu\varphi} +
\tilde{g}_{\mu\varphi} \tilde{g}_{\nu\beta} \Big] - {1\over 3}
\tilde{g}_{\mu\nu} \tilde{g}_{\beta\varphi} \Bigg\} \nnb \\
\cp \Bigg\{ - g_{\lambda\tau} + {p_\lambda^\prime p_\tau^\prime \over m_{V_Q}^2}
\Bigg\}~ \nnb \\
\es {1\over 2} g_2 f_T f_V m_{T_Q}^3 m_{V_Q} {1\over p^2-m_{T_Q}^2}
{1\over p^{\prime 2} - m_{V_Q}^2} \Bigg\{ \varepsilon^{\mu\alpha\eta\tau}
\Bigg[p_\alpha q_\nu q_\eta + p_\alpha p_\nu q_\eta
{m_{T_Q}^2 + m_{\cal P}^2 - m_{V_Q}^2\over 2 m_{T_Q}^2} \Bigg] \nnb \\
\ar (\mu \lrar \nu) \Bigg\}~,
\eea
where
\bea
\tilde{g}_{\mu\nu} = - g_{\mu\nu} + {p_\mu p_\nu \over m_{T_Q}^2}~.\nnb
\eea
We now turn our attention to the calculation of the correlation function
from the QCD side. i.e. in terms of the quark--gluon degrees of freedom.
After contracting the heavy quark fields the correlation function can be
written as,
\bea
\label{eozd13}
\Pi_{\mu\nu(\tau)} = -{1\over 2} \int d^4x e^{ipx} \lla {\cal P}(q) \vel
\bar{q}(x) \gamma_\mu \olra{\cal D}_\nu (x) S_Q (x) \gamma_{5(\tau)} q(0)
+ (\mu \lrar \nu) \ver 0 \rra~,
\eea
where $S_Q(x)$  is the heavy quark propagator whose expression in the
coordinate representation is given as,
\bea
\label{eozd14} 
S_Q(x) \es  {m_Q^2 \over 4 \pi^2} \Bigg\{ {K_1(m_Q\sqrt{-x^2}) \over
\sqrt{-x^2}} + i {\not\!{x} \over
-x^2} K_2(m_Q\sqrt{-x^2}) \Bigg\} \nnb \\
\ek {g_s m_Q \over 16 \pi^2} \int_0^1 du \Bigg[
G_{\mu\nu}(ux) \left(\sigma^{\mu\nu} \!\!\!\not\!{x} +
\not\!{x} \sigma^{\mu\nu}\right) {K_1 (m_Q\sqrt{-x^2})\over \sqrt{-x^2}}
+ 2 \sigma^{\mu\nu} K_0(m_Q\sqrt{-x^2})\Bigg]~,
\eea
where $K_n$ is the modified Bessel function. In further analysis we shall
use the following relations for the derivatives of the Bessel functions,
\bea
\label{eozd15}
{d\over dx} K_n (x) \es -{1\over 2} \Big[ K_{n-1} (x) + K_{n+1} (x) \Big]~,\nnb \\
{d\over dx} (x^n K_n ) \es - x^n K_{n-1}(x)~,\nnb \\
{d\over dx} (x^{-n} K_n ) \es - x^{-n} K_{n+1}(x)~.
\eea

It should be noted here that the expansion of the quark operator up to twist
four terms is calculated in \cite{Rozd18}, which gets contributions from
$\bar{q} G q$, $\bar{q} GG q$, and $\bar{q}q \bar{q}q$ nonlocal operators.
In the present work we consider operators with one gluon field. Neglecting
the contributions coming from the $\bar{q} GG q$ and $\bar{q}q
\bar{q}q$ operators can be justified on the basis of an expansion in
conformal spin.

Having the expression of the heavy quark operator at hand, we now calculate
the correlation function from the QCD side. The expression of the
correlation function in deep Eucledian domain $p_E^2 \to - \infty$,
$p^{\prime 2} \to - \infty$, can be obtained by using the operator product
expansion. It follows from Eq. (\ref{eozd13}) that in order to calculate the
correlation function from the QCD side the matrix elements of the nonlocal
operators between vacuum and one--particle light pseudoscalar meson states.
The matrix elements $\lla {\cal P}(q) \vel \bar{q} (x) \Gamma q^\prime (0) \ver 0
\rra$ are parametrized in terms of the distribution amplitudes (DAs)
\cite{Rozd19,Rozd20} as given below,
\bea
\label{eozd16} 
\lla {\cal P}(p)\vel \bar q(x) \gamma_\mu \gamma_5 q(0)\ver 0 \rra \es
-i f_{\cal P} q_\mu  \int_0^1 du  e^{i \bar u q x}
    \left( \varphi_{\cal P}(u) + {1\over 16} m_{\cal P}^2
x^2 {\Bbb{A}}(u) \right) \nnb \\
\ek {i\over 2} f_{\cal P} m_{\cal P}^2 {x_\mu\over qx}
\int_0^1 du e^{i \bar u qx} {\Bbb{B}}(u)~,\nnb \\
\lla {\cal P}(p)\vel \bar q(x) i \gamma_5 q(0)\ver 0 \rra \es
\mu_{\cal P} \int_0^1 du e^{i \bar u qx} \phi_P(u)~,\nnb \\
\lla {\cal P}(p)\vel \bar q(x) \sigma_{\alpha \beta} \gamma_5 q(0)\ver 0 \rra \es
{i\over 6} \mu_{\cal P} \left( 1 - \widetilde{\mu}_{\cal P}^2 \right)
\left( q_\alpha x_\beta - q_\beta x_\alpha\right)
\int_0^1 du e^{i \bar u qx} \phi_\sigma(u)~,\nnb \\
\lla {\cal P}(p)\vel \bar q(x) \sigma_{\mu \nu} \gamma_5 g_s
G_{\alpha \beta}(v x) q(0)\ver 0 \rra \es i \mu_{\cal P} \left[
q_\alpha q_\mu \left( g_{\nu \beta} - {1\over qx}(q_\nu x_\beta +
q_\beta x_\nu) \right) \right. \nnb \\
\ek q_\alpha q_\nu \left( g_{\mu \beta} -
{1\over qx}(q_\mu x_\beta + q_\beta x_\mu) \right) \nnb \\
\ek q_\beta q_\mu \left( g_{\nu \alpha} - {1\over qx}
(q_\nu x_\alpha + q_\alpha x_\nu) \right) \nnb \\
\ar q_\beta q_\nu \left. \left( g_{\mu \alpha} -
{1\over qx}(q_\mu x_\alpha + q_\alpha x_\mu) \right) \right] \nnb \\
\cp \int {\cal D} \alpha e^{i (\alpha_{\bar q} +
v \alpha_g) qx} {\cal T}(\alpha_i)~,\nnb \\
\lla {\cal P}(p)\vel \bar q(x) \gamma_\mu \gamma_5 g_s
G_{\alpha \beta} (v x) q(0)\ver 0 \rra \es q_\mu (q_\alpha x_\beta -
q_\beta x_\alpha) {1\over qx} f_{\cal P} m_{\cal P}^2
\int {\cal D}\alpha e^{i (\alpha_{\bar q} + v \alpha_g) qx}
{\cal A}_\parallel (\alpha_i) \nnb \\
\ar \left[q_\beta \left( g_{\mu \alpha} - {1\over qx}
(q_\mu x_\alpha + q_\alpha x_\mu) \right) \right. \nnb \\
\ek q_\alpha \left. \left(g_{\mu \beta}  - {1\over qx}
(q_\mu x_\beta + q_\beta x_\mu) \right) \right]
f_{\cal P} m_{\cal P}^2 \nnb \\
\cp \int {\cal D}\alpha e^{i (\alpha_{\bar q} + v \alpha _g)
q x} {\cal A}_\perp(\alpha_i)~,\nnb \\
\lla {\cal P}(p)\vel \bar q(x) \gamma_\mu i g_s G_{\alpha \beta}
(v x) q(0)\ver 0 \rra \es q_\mu (q_\alpha x_\beta - q_\beta x_\alpha)
{1\over qx} f_{\cal P} m_{\cal P}^2 \int {\cal D}\alpha e^{i (\alpha_{\bar q} +
v \alpha_g) qx} {\cal V}_\parallel (\alpha_i) \nnb \\
\ar \left[q_\beta \left( g_{\mu \alpha} - {1\over qx}
(q_\mu x_\alpha + q_\alpha x_\mu) \right) \right. \nnb \\
\ek q_\alpha \left. \left(g_{\mu \beta}  - {1\over qx}
(q_\mu x_\beta + q_\beta x_\mu) \right) \right] f_{\cal P} m_{\cal P}^2 \nnb \\
\cp \int {\cal D}\alpha e^{i (\alpha_{\bar q} +
v \alpha _g) q x} {\cal V}_\perp(\alpha_i)~,
\eea
where
\bea
\label{nolabel07}
\mu_{\cal P} = f_{\cal P} {m_{\cal P}^2\over m_{q_1} + m_{q_2}}~,~~~~~
\widetilde{\mu}_{\cal P} = {m_{q_1} + m_{q_2} \over m_{\cal P}}~, \nnb
\eea
and $q_1$ and $q_2$ are the quarks in the meson ${\cal P}$,
${\cal D}\alpha = d\alpha_{\bar q} d\alpha_q d\alpha_g
\delta(1-\alpha_{\bar q} - \alpha_q - \alpha_g)$, and
and the DA's $\Bbb{A}(u),$ $\Bbb{B}(u),$
$\varphi_{\cal P}(u),$ $\phi_P(u),$ $\phi_\sigma(u),$
${\cal T}(\alpha_i),$ ${\cal A}_\perp(\alpha_i),$ ${\cal A}_\parallel(\alpha_i),$
${\cal V}_\perp(\alpha_i)$ and ${\cal V}_\parallel(\alpha_i)$
are functions of definite twist and their expressions are given in
the next section.

Using Eq. (\ref{eozd16}) and corresponding coefficients of the Lorentz
structures the theoretical part of the correlation function can be
calculated straightforwardly. Equating the coefficients of the respective
Lorentz structures calculated from the hadronic and QCD sides of the
correlation function, and performing Borel transformation for the variables
$-p^2$ and $-p^{\prime 2}= -(p+q)^2$ in order to suppress the continuum and
higher states contribution, we obtain the sum rules for the strong coupling
constants of the heavy tensor with pseudoscalar heavy and light mesons given
as,
\bea
\label{nolabel17}
{f_{P_Q} m_{P_Q}^2 \over m_Q+m_q} f_T m_{T_Q}^3 g_1 A_i e^{-m_{T_Q}^2/M_1^2}
e^{-m_{P_Q}^2/M_2^2} \es \widetilde{\Pi}_i^{th.(P_Q)}~,\nnb \\
f_{V_Q} m_{V_Q} f_T m_{T_Q}^3 g_2 B_i e^{-m_{T_Q}^2/M_1^2}
e^{-m_{V_Q}^2/M_2^2} \es \widetilde{\Pi}_i^{th.(V_Q)}~,\nnb
\eea
where $\widetilde{ }$  means Borel transformed invariant function, $P_Q$ and
$V_Q$ refer to the heavy pseudoscalar and vector mesons, respectively. The expressions of
$A_i$ and $B_i$ in the above--expression are as follows,
\bea
\label{nolabel18}
A_i \es \left\{ \begin{array}{l}
 \phantom{-}1,~\mbox{for the $q_\mu q_\nu$ structure~,}\\ \nnb \\
\ds - {1\over 2 m_T^2} (m_{T_Q}^2 + m_{\cal P}^2 - m_{P_Q}^2),~\mbox{for the $q_\mu p_\nu +
q_\nu p_\mu $ structure~,}\\ \nnb \\              
-\ds{1\over 3} m_{\cal P}^2 + \ds{(m_{T_Q}^2+m_{\cal P}^2-m_{P_Q}^2)^2 \over 12
m_{T_Q}^2},~\mbox{for the $g_{\mu\nu}$ structure~,}\\ \nnb \\
\ds{1\over m_{T_Q}^4} \Bigg[\ds{1\over 6}(m_{T_Q}^2+m_{\cal P}^2-m_{P_Q}^2)^2 +
\ds{m_{\cal P}^2
m_{T_Q}^2 \over 3} \Bigg],~\mbox{for the $p_\mu p_\nu$ structure}~.
\end{array} \right. \nnb \\ \nnb \\ \nnb \\
\label{nolabel19}
B_i \es \left\{ \begin{array}{l}
 \phantom{-}1,~\mbox{for the $\varepsilon^{\mu\alpha\eta\tau} p_\alpha q_\nu
q_\eta$ $\left( \varepsilon^{\nu\alpha\eta\tau} p_\alpha q_\mu
q_\eta \right)$  structure~,}\\ \nnb \\
\ds{1\over 4 m_{T_Q}^2} (m_{T_Q}^2+m_{\cal P}^2-m_{P_Q}^2),~\mbox{for
the $\varepsilon^{\mu\alpha\eta\tau} p_\alpha p_\nu
q_\eta$ $\left( \varepsilon^{\nu\alpha\eta\tau} p_\alpha p_\mu q_\eta
\right)$ structure}~.
\end{array} \right.
\eea
The expressions of the $\widetilde{\Pi}_i^{th.(P_Q)}$ and
$\widetilde{\Pi}_i^{th.(V_Q)}$ are given as:
\\\\
% ...........................................................
{\bf 1) Coefficient of the $q_\mu q_\nu$ structure}
%
% (tmrest) rest of the terms
%
\bea
\widetilde{\Pi}_1^{th.(P)} \es e^{m_{\cal P}^2/4 M^2} e^{-m_Q^2/M^2} \Bigg\{
{1\over 48 M^2} f_{\cal P} m_{\cal P}^2 m_Q^3 \mathbb{A}(u_0)
% In[21]:= tmp1BB//InputForm
- {1\over 144} M^2 \Big\{12 f_{\cal P} m_Q \varphi_{\cal P}(u_0) \nnb \\ 
\ar    \mu_{\cal P} \Big[6 \varphi_{\cal P}(u_0) - (1 - \widetilde{\mu}_{\cal P}^2) \Big( 8 \phi_\sigma(u_0) + 
        \phi_\sigma^\prime(u_0)\Big) \Big] \Big\}
% In[25]:= tmrest//InputForm
+ {1\over 24} f_{\cal P} m_{\cal P}^2 m_Q \widetilde{j}_1(\mathbb{B}) \nnb \\
\ar {1\over 432 m_Q} \Big[ 9 f_{\cal P} m_{\cal P}^2 m_Q^2 \mathbb{A}(u_0) -
   6 m_Q (m_{\cal P}^2 - 2 m_Q^2) \mu_{\cal P} (1 - \widetilde{\mu}_{\cal P}^2)
\phi_\sigma(u_0)\Big]\Bigg\}\nnb ~.
\eea
\\\\
%%%%%%%%%%%%%
%%%%%%%%%%%%%
%%%%%%%%%%%%%
%%%%%%%%%%%%%
{\bf 2) Coefficient of the $p_\mu q_\nu + p_\nu q_\mu$ structure}
\bea
\widetilde{\Pi}_2^{th.(P)} \es e^{m_{\cal P}^2/4 M^2} e^{-m_Q^2/M^2} \Bigg\{
% In[7]:= tmm4//InputForm
   {1\over 48 M^2} f_{\cal P} m_{\cal P}^2 m_Q^3 \mathbb{A}(u_0)
% In[16]:= tmp1BB//InputForm
- {1\over 72} M^2 \Big\{ 6 f_{\cal P} m_Q \varphi_{\cal P}(u_0) \nnb \\
\ar \mu_{\cal P}
\Big[6 \varphi_{\cal P}(u_0) - (1 - \widetilde{\mu}_{\cal P}^2) 
       (4 \phi_\sigma(u_0) + \phi_\sigma^\prime(u_0)\Big)\Big]\Big\}
% In[21]:= tmrest//InputForm
+ {1\over 12} f_{\cal P} m_{\cal P}^2 m_Q \widetilde{j}_1(\mathbb{B}) \nnb \\
\ar {1\over 432 m_Q} \Big[9 f_{\cal P} m_{\cal P}^2 m_Q^2 \mathbb{A}(u_0) -
   12 m_Q (m_{\cal P}^2 - m_Q^2) \mu_{\cal P} (1 - \widetilde{\mu}_{\cal P}^2)
\phi_\sigma(u_0)\Big] \Bigg\}\nnb ~.
\eea
\\\\
%%%%%%%%%%%%%
%%%%%%%%%%%%%
%%%%%%%%%%%%%
%%%%%%%%%%%%%
{\bf 3) Coefficient of the $g_{\mu\nu}$ structure}
\bea
\widetilde{\Pi}_3^{th.(P)} \es e^{m_{\cal P}^2/4 M^2} e^{-m_Q^2/M^2} \Bigg\{
% In[12]:= tmp2//InputForm
 - {1\over 72} M^4 \mu_{\cal P} \Big[12 \varphi_{\cal P}(u_0) - (1 - \widetilde{\mu}_{\cal P}^2)
\phi_\sigma^\prime(u_0)\Big] \nnb \\
% In[13]:= tmp1BB//InputForm
\ar {1\over 12} M^2 f_{\cal P} m_{\cal P}^2 m_Q \widetilde{j}_1(\mathbb{B})
-  {1\over 36} M^2 m_{\cal P}^2 \mu_{\cal P} (1 - \widetilde{\mu}_{\cal P}^2)
\phi_\sigma(u_0) \Bigg\}~.
\eea
\\\\
%%%%%%%%%%%%%
%%%%%%%%%%%%%
%%%%%%%%%%%%%
%%%%%%%%%%%%%
{\bf 4) Coefficient of the $p_\mu p_\nu$ structure}
\bea
\widetilde{\Pi}_4^{th.(P)} \es e^{m_{\cal P}^2/4 M^2} e^{-m_Q^2/M^2} \Bigg\{
% In[13]:= tmp1BB//InputForm
 - {1\over 36} M^2 \mu_{\cal P} \Big[6 \varphi_{\cal P}(u_0) -
(1 - \widetilde{\mu}_{\cal P}^2) \phi_\sigma^\prime(u_0)\Big] \nnb \\
% In[17]:= tmrest//InputForm
\ar {1\over 6} f_{\cal P} m_{\cal P}^2 m_Q \widetilde{j}_1(\mathbb{B})
- {1\over 18} m_{\cal P}^2 \mu_{\cal P} (1 - \widetilde{\mu}_{\cal P}^2) \phi_\sigma(u_0)\nnb
\Bigg\} \nnb ~.
\eea
\\\\
%%%%%%%%%%%%%
%%%%%%%%%%%%%
%%%%%%%%%%%%%
%%%%%%%%%%%%%
{\bf 5) Coefficient of the 
$\varepsilon^{\mu\alpha\eta\tau} p_\alpha q_\nu q_\eta$ and
$\varepsilon^{\nu\alpha\eta\tau} p_\alpha q_\mu q_\eta$
structures}
\bea
\widetilde{\Pi}_1^{th.(V)} \es e^{m_{\cal P}^2/4 M^2} e^{-m_Q^2/M^2} \Bigg\{
% In[8]:= tmm1//InputForm
 - {1\over 96 m_Q M^2} f_{\cal P} m_{\cal P}^2 m_Q^3 \mathbb{A}(u_0) \nnb \\
% In[13]:= tmp1BB//InputForm
\ar {1\over 24} f_{\cal P} M^2 \varphi_{\cal P}(u_0)
% In[18]:= tmrest//InputForm
- {1\over 144} \Big[3 f_{\cal P} m_{\cal P}^2 \mathbb{A}(u_0) + 2 m_Q \mu_{\cal P}
(1 - \widetilde{\mu}_{\cal P}^2) \phi_\sigma(u_0)\Big]\nnb \Bigg\} \nnb ~.
\eea
\\\\
%%%%%%%%%%%%%
%%%%%%%%%%%%%
%%%%%%%%%%%%%
%%%%%%%%%%%%%
{\bf 6) Coefficient of the 
$\varepsilon^{\mu\alpha\eta\tau} p_\alpha p_\nu q_\eta$ and
$\varepsilon^{\nu\alpha\eta\tau} p_\alpha p_\mu q_\eta$ 
structures}
\bea
\widetilde{\Pi}_2^{th.(V)} \es e^{m_{\cal P}^2/4 M^2} e^{-m_Q^2/M^2} \Bigg\{
% In[8]:= tmm1//InputForm
 - {1\over 48 m_Q M^2} f_{\cal P} m_{\cal P}^2 m_Q^3 \mathbb{A}(u_0) \nnb \\
% In[12]:= tmp1BB//InputForm
\ar {1\over 12} f_{\cal P} M^2 \varphi_{\cal P}(u_0)
% In[16]:= tmrest//InputForm
- {1\over 72} \Big[3 f_{\cal P} m_{\cal P}^2 \mathbb{A}(u_0) + 2 m_Q \mu_{\cal P}
(1 - \widetilde{\mu}_{\cal P}^2) \phi_\sigma(u_0)\Big]\nnb \nnb \Bigg\}~,
\eea         
%%%%%%%%%%%%%
where 
\bea
u_0={M_1^2 \over M_1^2
+M_2^2}~,~~~~~M^2={M_1^2 M_2^2 \over M_1^2 +M_2^2}\nnb ~.
\eea

The function $\widetilde{j}_1(f(u))$ appearing in the invariant
functions above 
%$\widetilde{j}_2(f(u))$
is defined as:
\bea
\label{nolabel}
\widetilde{j}_1(f(u)) \es \int_{u_0}^1 du f(u)~. \nnb \\
\eea

It should be noted here that, in the above expressions the light quark
masses $m_u$, $m_d$ and $m_s$ are all set to zero, while in the numerical
calculations the mass $m_s$ of the strange quark is taken into account.
It should also be mentioned that the derivation of the double
spectral density for the higher twist contributions is not trivial, and
it is calculated in \cite{Rozd21}. In our calculations the higher twist
contributions appear in the similar form as they do in \cite{Rozd21}.
Therefore, for more detail about this issue, the interested reader
can consult this work.
    
\section{Numerical analysis}
This section is devoted to the numerical analysis of the sum rules for the
heavy tensor mesons with the heavy pseudoscalar (vector) and light
pseudoscalar mesons.

The main input parameters of the light cone QCD sum rules are the
distribution amplitudes (DAs), whose expressions are given below
\cite{Rozd19,Rozd20},

\bea
\label{eozd17}
\varphi_{\cal P}(u) \es 6 u \bar u \left[ 1 + a_1^{\cal P} C_1(2 u -1) +
a_2^{\cal P} C_2^{3/2}(2 u - 1) \right]~,  \nnb \\
{\cal T}(\alpha_i) \es 360 \eta_3 \alpha_{\bar q} \alpha_q
\alpha_g^2 \left[ 1 + w_3 {1\over 2} (7 \alpha_g-3) \right]~, \nnb \\
\phi_P(u) \es 1 + \left[ 30 \eta_3 - {5\over 2}
{1\over \mu_{\cal P}^2}\right] C_2^{1/2}(2 u - 1)~,  \nnb \\
\ar \left( -3 \eta_3 w_3  - {27\over 20} {1\over \mu_{\cal P}^2} -
{81\over 10} {1\over \mu_{\cal P}^2} a_2^{\cal P} \right)
C_4^{1/2}(2u-1)~, \nnb \\
\phi_\sigma(u) \es 6 u \bar u \left[ 1 + \left(5 \eta_3 - {1\over 2} \eta_3 w_3 -
{7\over 20}  \mu_{\cal P}^2 - {3\over 5} \mu_{\cal P}^2 a_2^{\cal P} \right)
C_2^{3/2}(2u-1) \right] ~, \nnb \\
{\cal V}_\parallel(\alpha_i) \es 120 \alpha_q \alpha_{\bar q} \alpha_g
\left( v_{00} + v_{10} (3 \alpha_g -1) \right) ~, \nnb \\
{\cal A}_\parallel(\alpha_i) \es 120 \alpha_q \alpha_{\bar q} \alpha_g
\left( 0 + a_{10} (\alpha_q - \alpha_{\bar q}) \right) ~, \nnb \\
{\cal V}_\perp (\alpha_i) \es - 30 \alpha_g^2\left[ h_{00}(1-\alpha_g) +
h_{01} (\alpha_g(1-\alpha_g)- 6 \alpha_q \alpha_{\bar q}) +
h_{10}(\alpha_g(1-\alpha_g) - {3\over 2} (\alpha_{\bar q}^2+
\alpha_q^2)) \right] ~, \nnb \\
{\cal A}_\perp (\alpha_i) \es 30 \alpha_g^2(\alpha_{\bar q} - \alpha_q)
\left[ h_{00} + h_{01} \alpha_g + {1\over 2} h_{10}(5 \alpha_g-3) \right] ~, \nnb \\
B(u)\es g_{\cal P}(u) - \varphi_{\cal P}(u) ~, \nnb \\
g_{\cal P}(u) \es g_0 C_0^{1/2}(2 u - 1) + g_2 C_2^{1/2}(2 u - 1) +
g_4 C_4^{1/2}(2 u - 1) ~, \nnb \\
\Bbb{A}(u) \es 6 u \bar u \left[{16\over 15} + {24\over 35} a_2^{\cal P}+
20 \eta_3 + {20\over 9} \eta_4 +
\left( - {1\over 15}+ {1\over 16}- {7\over 27}\eta_3 w_3 -
{10\over 27} \eta_4 \right) C_2^{3/2}(2 u - 1)  \right. \nnb \\
    \ar \left. \left( - {11\over 210}a_2^{\cal P} - {4\over 135}
\eta_3w_3 \right)C_4^{3/2}(2 u - 1)\right] ~, \nnb \\
\ar \left( -{18\over 5} a_2^{\cal P} + 21 \eta_4 w_4 \right)
\left[ 2 u^3 (10 - 15 u + 6 u^2) \ln u  \right. \nnb \\
\ar \left. 2 \bar u^3 (10 - 15 \bar u + 6 \bar u ^2) \ln\bar u +
u \bar u (2 + 13 u \bar u) \right]~,
\eea
where $C_n^k(x)$ are the Gegenbauer polynomials, and
\bea
\label{eozd18}
h_{00}\es v_{00} = - {1\over 3}\eta_4 ~, \nnb \\
a_{10} \es {21\over 8} \eta_4 w_4 - {9\over 20} a_2^{\cal P} ~, \nnb \\
v_{10} \es {21\over 8} \eta_4 w_4 ~, \nnb \\
h_{01} \es {7\over 4}  \eta_4 w_4  - {3\over 20} a_2^{\cal P} ~, \nnb \\
h_{10} \es {7\over 4} \eta_4 w_4 + {3\over 20} a_2^{\cal P} ~, \nnb \\
g_0 \es 1 ~, \nnb \\
g_2 \es 1 + {18\over 7} a_2^{\cal P} + 60 \eta_3  + {20\over 3} \eta_4 ~, \nnb \\
g_4 \es  - {9\over 28} a_2^{\cal P} - 6 \eta_3 w_3~.
\eea
The values of the parameters $a_1^{\cal P}$, $a_2^{\cal P}$,
$\eta_3$, $\eta_4$, $w_3$, and $w_4$ entering Eq. (\ref{eozd18}) are listed in
Table (\ref{param}) for the pseudoscalar $\pi$, $K$ and $\eta$ mesons.

\begin{table}[h]
\def\bos{\lower 0.25cm\hbox{{\vrule width 0pt height 0.7cm}}}
\begin{center}
\begin{tabular}{|c|c|c|}
\hline\hline
        & \bos  $\pi$   &   $K$ \\
\hline
$a_1^{\cal P}$  & \bos   0 &   0.050 \\
\hline
$a_2^{\cal P}~\mbox{(set-1)}$  & \bos   0.11  &   0.15 \\
\hline
$a_2^{\cal P}~\mbox{(set-2)}$  &  \bos  0.25  &   0.27 \\
\hline
$\eta_3$    & \bos  0.015 &   0.015 \\
\hline
$\eta_4$    & \bos  10    &   0.6 \\
\hline
$w_3$       & \bos  $-3$    &   $-3$ \\
\hline
$w_4$       & \bos  0.2   &   0.2 \\
\hline \hline
\end{tabular}
\end{center}
\caption{Parameters of the wave function calculated at the renormalization scale $\mu = 1 ~GeV$}
\label{param}
\end{table}

In addition to the above--mentioned parameters, the decay constants of the
heavy tensor, heavy pseudoscalar (vector) and light pseudoscalar mesons, and
masses of the quarks. The decay constants of the heavy tensor mesons are
calculated in \cite{Rozd16,Rozd17,Rozd22} to have the values $f_{{\cal D}_2^\ast}=
(0.018 \pm 0.007)$, $f_{{\cal D}_{S_2}^\ast}=(0.023\pm 0.011)$,
$f_{{\cal B}_2^\ast}=0.011$, $f_{{\cal B}_{S_2}^\ast}=0.013$,
$f_{{\cal D}^\ast}=0.24~GeV$, $f_{{\cal B}^\ast}=0.16~GeV$
\cite{Rozd21}.

In the present work we use the $\overline{MS}$ values of the 
quark masses predicted by the particle data group whose values are
\cite{Rozd08}: $\overline{m}_c(m_c)= (1.275 \pm 0.025)~GeV$,
$\overline{m}_b(m_b)= (4.18 \pm 0.03)~GeV$, and $m_s(2~GeV)= (0.095 \pm
0.005)~GeV$. The masses of the heavy mesons we use in the present work
are calculated in framework of the the QCD sum rules, having the values,
$m_{{\cal D}_2^\ast}=(2.460\pm 0.009)~GeV$, $m_{{\cal B}_2^\ast}=(5.73\pm 0.06)~GeV$, $m_{{\cal
B}_{S_2}^\ast}=(5.84\pm 0.06)~GeV$, which are all very close to their
experimental values.

There are two extra parameters entering to the sum rules, namely, the
continuum threshold $s_0$ and the Borel mass parameters $M_1^2$ and $M_2^2$.
In the present analysis we set $M_1^2 = M_2^2 = 2 M^2$, and this choice 
corresponds to $u_0=1/2$. The continuum threshold $s_0$ is determined from
an analysis of the two--point correlation function which leads to the
following results: $s_{0,{\cal D}_2^\ast}=(8.5\pm 0.5)~GeV^2$,
$s_{0,{\cal D}_{S_2}^\ast}=(9.5\pm 0.5)~GeV^2$,
$s_{0,{\cal B}_2^\ast}=(39.0\pm 1.0)~GeV^2$,
$s_{0,{\cal B}_{S_2}^\ast}=(41.0\pm 1.0)~GeV^2$ \cite{Rozd16,Rozd17}.

Of course, we need to find such a region of $M^2$ where the strong coupling
constants $g_{T_Q P_Q {\cal P}}$ and $g_{T_Q V_Q {\cal P}}$ are insensitive to the
variation in $M^2$. The upper bound of $M^2$is determined from the condition
that the higher states and continuum contributions constitute, say, 30\% of
the total result. The lower bound of the Borel mass parameter $M^2$ is
determined from the condition that the contribution of the highest term
with the power $1/M^2$ is less than 25\% of the contribution coming from
from the highest power of $M^2$. These two conditions lead to the following
working regions of $M^2$:
$2~GeV^2 \le M^2 \le 4~GeV^2$ (for ${\cal D}_2^\ast$ and ${\cal
D}_{S_2}^\ast$),
$4~GeV^2 \le M^2 \le 7~GeV^2$ (for ${\cal B}_2^\ast$ and ${\cal
B}_{S_2}^\ast$). 

Having decided the working regions $M^2$, we now calculate the strong
coupling constants for the $T_Q P_Q {\cal P}$ and $T_Q V_Q {\cal P}$ vertices.
Performing similar analysis we obtain the values of the coupling constants
for the $T_Q P_Q (V_Q) {\cal P}$ vertices which are presented in Tables 2 and 3.
In these calculations we use two different sets of parameters that appear in
the expressions of the DAs.

% .........................................................

\begin{table}[h]
\centering
\renewcommand{\arraystretch}{1.3}
\addtolength{\arraycolsep}{-0.5pt}
\footnotesize
$$
\begin{array}{|l|c|c||c|c||c|c|c|c|c|c|c|c|}
\hline \hline  
  \multirow{3}{*}{ }        &\multicolumn{2}{c||}{\cite{Rozd15}}   
                            &\multicolumn{2}{c||}{\cite{Rozd16}}
                            &\multicolumn{8}{c|}{\mbox{present work}}
\\ \cline{2-13}
	      &   \multicolumn{1}{c|}{\multirow{2}{*}{\mbox{$g_{\mu\nu}$}}}   &
                  \multicolumn{1}{c||}{\multirow{2}{*}{\mbox{$p_\mu p_\nu$}}} &
                  \multicolumn{1}{c|}{\multirow{2}{*}{\mbox{$g_{\mu\nu}$}}}   & 
                  \multicolumn{1}{c||}{\multirow{2}{*}{\mbox{$p_\mu p_\nu$}}} &
              \multicolumn{2}{c|}{\mbox{$g_{\mu\nu}$}}   &
              \multicolumn{2}{c|}{\mbox{$p_\mu p_\nu$}}  &
              \multicolumn{2}{c|}{\mbox{$q_\mu q_\nu$}}  &
              \multicolumn{2}{c|}{\mbox{$p_\mu q_\nu + p_\nu q_\mu$}}
\\ \cline{6-13}
             & \multicolumn{1}{c|}{}
             & \multicolumn{1}{c||}{}  
             & \multicolumn{1}{c|}{}  
             & \multicolumn{1}{c||}{}  
             &     \multicolumn{1}{c|}{\mbox{set-1}}    &
                   \multicolumn{1}{c|}{\mbox{set-2}}    &
                   \multicolumn{1}{c|}{\mbox{set-1}}    &
                   \multicolumn{1}{c|}{\mbox{set-2}}    &
                   \multicolumn{1}{c|}{\mbox{set-1}}    &
                   \multicolumn{1}{c|}{\mbox{set-2}}    &
                   \multicolumn{1}{c|}{\mbox{set-1}}    &
                   \multicolumn{1}{c|}{\mbox{set-2}}     \\ \hline
 {\cal D} \to {\cal D} \pi^+        & 15.3 & 4.63 & 16.5  & 12.3 & 39 \pm 13 &
40 \pm 3  & 11  \pm 1   &12 \pm 1 & 0.10  & 0.08 & 17 \pm 5    & 17 \pm 6  \\
 {\cal D}_{S_2} \to {\cal D} K^0    & 18.3 & 5.76  & 12.2 & 9.9 & 78 \pm 20   &
74 \pm 5  & 10  \pm 1   &10 \pm 1 & 0.10  & 0.13 & 19 \pm 6    & 9 \pm 3    \\
 {\cal B} \to {\cal B} \pi^+        & -  & -  & 39.3     & 17.1    & 90 \pm 25     &
88 \pm 20        & 70 \pm 15         &73  \pm 20    & 0.30  & 0.15 & 40 \pm 12 & 36 \pm 12 \\
 {\cal B}_{S_2} \to {\cal B} K^0    & -  & -  & 26.3     & 12.9    & 400 \pm 100     &
360 \pm50       & 58 \pm 12        &55 \pm 12      & 0.25  & 0.30 & 24 \pm 8 & 22 \pm
8 \\
\hline \hline
\end{array}
$$

\caption{The values of the strong coupling constants for the $T_Q P_Q {\cal P}$
vertices for different Lorentz structures}

\renewcommand{\arraystretch}{1}
\addtolength{\arraycolsep}{-1.0pt}

\end{table}

% .........................................................
% .........................................................

\begin{table}[h]
\centering
\renewcommand{\arraystretch}{1.3}
\addtolength{\arraycolsep}{-0.5pt}
\small
$$
\begin{array}{|l|c|c||c|c|c|c|}
\hline \hline  
  \multirow{3}{*}{ }        &\multicolumn{2}{c||}{\cite{Rozd16}}   
                            &\multicolumn{4}{c|}{\mbox{present work}}
\\ \cline{2-7} &
\multicolumn{1}{c|}{\multirow{2}{*}{\mbox{$\varepsilon_{\nu\tau\rho\sigma}
p^{\prime\mu} p^\rho p^{\prime\sigma}$}}} &     
\multicolumn{1}{c||}{\multirow{2}{*}{\mbox{$\varepsilon_{\nu\tau\rho\sigma}
p^\mu p^\rho p^{\prime\sigma}$}}}  &
\multicolumn{2}{c|}{\mbox{$ \varepsilon_{\rho \sigma \alpha \tau}
p^\beta p^\rho q^\sigma $}} &
\multicolumn{2}{c|}{\mbox{$\varepsilon_{\rho \sigma \alpha \tau}
q^\beta p^\rho q^\sigma$}}
\\ \cline{4-7} &
             \multicolumn{1}{c|}{}  &
             \multicolumn{1}{c||}{} & 
             \multicolumn{1}{c|}{\mbox{set-1}} &
             \multicolumn{1}{c|}{\mbox{set-2}} &
             \multicolumn{1}{c|}{\mbox{set-1}} &
             \multicolumn{1}{c|}{\mbox{set-2}}
\\ \hline
 {\cal D}_2^0 \to {\cal D}^{\ast +} \pi^-    & 4.00 & 0.73  & 4.4 \pm 1.0 
     & \mbox{no stability}            & \phantom{-}0.42  \pm 0.10
     & \mbox{no stability}  \\
 {\cal D}_{S_2}^0 \to {\cal D}^{\ast -} K^+    & 2.98 & 0.79  & 4.4 \pm 1.0
     & \mbox{no stability}            & -0.37 \pm 0.13   & \mbox{no stability}  \\
 {\cal B}_2^0 \to {\cal B}^{\ast +} \pi^-      & 3.87 &  -    & 5.9 \pm 2.0
     & \mbox{no stability}      & \phantom{-}0.22 \pm 0.05     & \mbox{no stability}   \\
 {\cal B}_{S_2}^0 \to {\cal B}^{\ast +} K^-    & 2.89 &  -    & 3.0 \pm 1.0
     & \mbox{no stability}      & \phantom{-}0.28 \pm 0.06      & \mbox{no stability}  \\
\hline \hline
\end{array}
$$

\caption{Same as in Table 2, but for the  $T_Q V_Q {\cal P}$ vertices}

\renewcommand{\arraystretch}{1}
\addtolength{\arraycolsep}{-1.0pt}

\end{table}

% .........................................................

In Table 3 ``no stability" means that  there is no region of the Borel
parameter $M^2$ where the results for the strong coupling constants are
insensitive to its variation.

From these Tables we deduce the following conclusions:

\begin{itemize}

\item The value of the strong coupling constant depends very strongly on the
choice of the corresponding Lorentz structure, especially for the $T_Q \to
P_Q {\cal P}$ transition.

\item From our analysis we observe that the value of the strong coupling
constant for the $T_Q \to P_Q {\cal P}$ transition ranges in a rather wide region
from $0.3(0.15)$ to $290(400)$ for the ${\cal B}_2 \to {\cal B}^- \pi^+$ $({\cal
B}_{S_2}^\ast \to {\cal B}^0 K^0)$. In the case of ${\cal D}_{2(S_2)} \to
{\cal D}^+ \pi^-\,({\cal D}^0 K^0)$ transitions the values of the
corresponding coupling constants vary in the range $0.15$ to $78.0$. These
results also show that our analysis that the values of the strong
coupling constants are very sensitive to the values of the parameters
appearing in the DAs.         

\end{itemize}

It follows from Table 1 that the most reliable value for the $T_Q P_Q {\cal P}$
vertex follows from the $q_\mu p_\nu + q_\nu p_\mu$ structure, from which we
get,
\bea
\label{eozd19}
g_1 \es \left\{ \begin{array}{l}
17 \pm 5,~\mbox{${\cal D}_2^0 \to {\cal D}^+ \pi^-$}~,\\
19 \pm 6,~\mbox{${\cal D}_{S_2}^+ \to {\cal D}^0 K^+$}~,\\
40 \pm 12,~\mbox{${\cal B}_2^0 \to {\cal B}^- \pi^+$}~,\\
24 \pm 8,~\mbox{${\cal B}_{S_2}^0 \to {\cal B}^0 K^0$}~.
\end{array} \right.
\eea

The strong coupling constants for the  $T_Q \to V_Q {\cal P}$ transitions is
obtained for the $\varepsilon_{\rho\sigma\alpha\tau} p^\beta p^\rho q^\sigma$
structure, and for the Set 1 values of the parameters of the wave functions
are used. As the result, for the coupling constants of the $T_Q V_Q {\cal P}$
vertex we get,
\bea
\label{Rozd20}
g_2 \es \left\{ \begin{array}{l}
4.4 \pm 1.0,~\mbox{${\cal D}_2^0 \to {\cal D}^{\ast +} \pi^-$}~,\\
4.4 \pm 1.0,~\mbox{${\cal D}_{S_2}^+ \to {\cal D}^{\ast 0} K^+$}~,\\
5.9 \pm 2.0,~\mbox{${\cal B}_2^0 \to {\cal B}^{\ast -} \pi^+$}~,\\
3.0 \pm 1.0,~\mbox{${\cal B}_{S_2}^0 \to {\cal B}^{\ast +} K^-$}~.
\end{array} \right.
\eea

We now compare our results on the strong coupling constants with those
predicted by the 3--point QCD sum rules method. For the structure
$g_{\mu\nu}$ our prediction for the coupling constant $g_{{\cal D}_2^\ast
{\cal D}^+ \pi^-}$ is approximately two times larger compared to the one
predicted in \cite{Rozd15} and \cite{Rozd16}. For the ${\cal B}_2 \to {\cal
B}^- \pi^+$ and ${\cal B}_{S_2} \to {\cal B}^0 K^0$ transition coupling
constants our results are two and more than ten times larger than the ones predicted
in \cite{Rozd16}.

At the end of this section we present the results for the decay widths of
all above--considered transitions, whose expressions are gives as,
\bea
\label{Rozd21}
\Gamma_{T_Q P_Q {\cal P}} \es {g_1^2 \vel \vec{p} \ver^5 \over 60 \pi m_{T_Q}^2}~, \nnb \\
\Gamma_{T_Q V_Q {\cal P}} \es {g_2^2 \vel \vec{p} \ver^5 \over 40 \pi}~,
\eea
where
\bea
\label{Rozd22}
\vel \vec{p} \ver = {1\over 2 m_{T_Q}} \left( m_{T_Q}^4 +m_{P_Q}^4 + m_{\cal P}^4
- 2 m_{T_Q}^2 m_{P_Q}^2 - 2 m_{T_Q}^2 m_{\cal P}^2 - 2 m_{P_Q}^2
m_{\cal P}^2\right)^{1/2}~.
\eea
Note that the $\vel \vec{p} \ver^5$ dependence is an indication of the fact
that the decay takes place at the D-wave level. The results for the decay
widths can be summarized as follows.

\begin{itemize}

\item In the case for the transitions of the heavy tensor mesons to heavy pseudoscalar
and light pseudoscalar mesons (for the $p_\mu q_\nu + p_\nu q_\mu$ structure),

\bea
\label{Rozd23}
\Gamma\left({\cal D}_2^0 (2460) \to {\cal D}^+ \pi^- \right) \es 8.6 \times 10^{-3}~GeV \nnb~, \\
\Gamma\left({\cal D}_{S_2}^+ (2573) \to {\cal D}^0 K^0 \right) \es 4.4 \times 10^{-3}~GeV \nnb~, \\
\Gamma\left({\cal B}_2^0 (5747) \to {\cal B}^- \pi^+ \right) \es 3.7 \times 10^{-3}~GeV \nnb~, \\
\Gamma\left({\cal B}_{S_2}^0 (5840) \to {\cal B}^0 K^0 \right) \es 8.6 \times10^{-5}~GeV~.
\eea

\item In the case for the transitions of the heavy tensor mesons to heavy
vector and light pseudoscalar mesons (for the
$\varepsilon_{\rho\sigma\alpha\tau} p^\beta p^\rho q^\sigma$ structure),

\bea
\label{Rozd24}
\Gamma\left({\cal D}_2^0 (2460) \to {\cal D}^{\ast +} \pi^- \right) \es 5.2 \times 10^{-3}~GeV \nnb~, \\
\Gamma\left({\cal D}_{S_2}^+ (2573) \to {\cal D}^{\ast 0} K^+ \right) \es 2.3 \times 10^{-3}~GeV \nnb~, \\
\Gamma\left({\cal B}_2^0 (5747) \to {\cal B}^{\ast -} \pi^+ \right) \es 4.0 \times 10^{-3}~GeV \nnb~, \\
§\Gamma\left({\cal B}_{S_2}^0 (5840) \to {\cal B}^{\ast 0} K^0 \right] \es 6.9 \times 10^{-5}~GeV \nnb~.  
\eea

\end{itemize}

It follows from the experimental data that \cite{Rozd08},
\bea
\label{nolabel31}
{\Gamma({\cal D}_2 (2460) \to {\cal D}^+ \pi^-) \over
 \Gamma({\cal D}_2 (2460) \to {\cal D}^{\ast +} \pi^-)} \es 1.55~, \nnb\\
{\Gamma({\cal B}_2 (5747) \to {\cal B}^+ \pi^-) \over
 \Gamma({\cal B}_2 (5747) \to {\cal B}^{\ast +} \pi^-)} \es 0.91~,\nnb
\eea
as well as from the BaBar Collaboration data \cite{Rozd23}
\bea
\label{nolabel32}
{\Gamma({\cal D}_2 (2460) \to {\cal D}^+ \pi^-) \over
 \Gamma({\cal D}_2 (2460) \to {\cal D}^+ \pi^-) +
 \Gamma({\cal D}_2 (2460) \to {\cal D}^{\ast +} \pi^-)} = 0.62 \pm 0.03 \pm
0.02~.\nnb
\eea
When we calculate the same ratios from Eqs. (\ref{Rozd23}) and
(\ref{Rozd24}), we obtain that,
\bea
\label{nolabel33}
{\Gamma({\cal D}_2 (2460) \to {\cal D}^+ \pi^-) \over
 \Gamma({\cal D}_2 (2460) \to {\cal D}^{\ast +} \pi^-)} \es 1.64~, \nnb\\
{\Gamma({\cal B}_2 (5747) \to {\cal B}^- \pi^+) \over
 \Gamma({\cal B}_2 (5747) \to {\cal B}^{\ast -} \pi^+)} \es 0.93~, \nnb
\eea
and
\bea
\label{nolabel34}  
{\Gamma({\cal D}_2 (2460) \to {\cal D}^+ \pi^-) \over
 \Gamma({\cal D}_2 (2460) \to {\cal D}^+ \pi^-) +
 \Gamma({\cal D}_2 (2460) \to {\cal D}^{\ast +} \pi^-)} = 0.67~. \nnb
\eea
We see that our prediction on the ratio of the decay widths are in good
agreement with the experimental results, as well as they are quite close to
the results of the works \cite{Rozd16} and \cite{Rozd17}.

Few words about the perturbative ${\cal O}(\alpha_s)$ corrections are in
order. These corrections increase the correlation function of the coupling constant of the
${\cal B}^\ast \to {\cal B} \pi$ transition about 50\% in the light cone QCD
sum rules \cite{Rozd24}. If we assume that this increase in the correlation
function is correct, then the we expect that the coupling constant increases
at the same order, i.e., $g_i \to 1.5 g_i$. This increase in the coupling
constant doubles the values of the decay widths as well.  

In conclusion, we calculate the strong coupling constant of the heavy
tensor meson to the heavy pseudoscalar (vector) and light pseudoscalar
mesons. It is seen that the values of the hadronic decay constants
are very strongly dependent on the choice of the Lorentz structure.
Furthermore, using the value of the coupling constants calculated in this
work we also estimate the corresponding decay widths. A comparison of our
predictions on these hadronic coupling constants with the results of the
3--point sum rules is presented.

% ...........................................................

\end{document}